\documentclass[a4paper,fleqn,usenatbib]{mnras}
\usepackage{newtxtext,newtxmath}
\usepackage[T1]{fontenc}

\DeclareRobustCommand{\VAN}[3]{#2}
\let\VANthebibliography\thebibliography
\def\thebibliography{\DeclareRobustCommand{\VAN}[3]{##3}\VANthebibliography}

\usepackage{graphicx}	
\usepackage{amsmath}	
\usepackage{natbib}

\defcitealias{can11}{C11}
\defcitealias{can13}{C13}

\usepackage[dvipsnames]{xcolor}

\newcommand{\vecF}{\mbox{\boldmath $F$} {}}

\title{Bondi-Hoyle-Lyttleton accretion flow in a stratified layer}

\author{
  F. J. S\'anchez-Salcedo\altaffilmark{1}}

\author[F. J. S\'anchez-Salcedo]{F. J. S\'anchez-Salcedo\thanks{E-mail: jsanchez@astro.unam.mx} \\
Universidad Nacional Aut\'onoma de M\'exico, Instituto de Astronom\'\i a, 
P.O. Box 70-264, Ciudad Universitaria, 04510, Mexico City, Mexico\
}

\date{Accepted XXX. Received YYY; in original form ZZZ}

\pubyear{2024}

\begin{document}
\label{firstpage}
\pagerange{\pageref{firstpage}--\pageref{lastpage}}
\maketitle

\begin{abstract}

We compute the density and velocity profiles along the tail induced by a 
body of mass $M$, embedded in the midplane of a vertically-stratified media with scaleheight $H$, adopting a one-dimensional model as in the Bondi-Hoyle-Lyttleton
problem. In analogy to what occurs in the case of a homogeneous medium, there exist 
a family of solutions that satisfy the boundary conditions. A shooting method is
employed to isolate those solutions that fulfill a specific set of physical and
mathematical constraints. The tail is found to be both densest and slowest when the
scaleheight $H$ is equal to the gravitational radius 
$\xi_{0}\equiv GM/v_{0}^{2}$, where $v_{0}$ its relative velocity with respect to the
medium. The location of the stagnation point is evaluated as a function of $H$ and
$\xi_{0}$, and an empirical fitting formula is provided. 
While the distance to the stagnation point is maximized when $H\simeq \xi_{0}$,
the mass accretion rate attains its maximum value for $H \ll \xi_{0}$ at fixed surface 
density. When instead the midplane density is held constant
and $H$ is varied, the accretion rate hardly changes once
$H$ exceeds about $2\xi_{0}$. Additionally, 
we investigate how both the drag force resulting from mass accretion and the gravitational
drag arising from its tail depend on $H/\xi_{0}$. We highlight how the effect of varying 
the degree of mixing in the tail influences the resulting drag force.
Finally, for the particular case of an infinitely
thin layer, we provide a simple analytical solution, which may serve as
a useful pedagogical reference.

\end{abstract}


\section{Introduction}

Gas accretion onto a gravitating object remains an active area of research in astrophysics
with applications including galaxy formation, star formation, compact objects 
and stellar binary systems \citep[see][for a review]{edg04}. 
The steady-state model for spherical accretion proposed by \citet{bon52} and the
Bondi, Hoyle and Lyttleton (BHL) description for accretion onto moving objects \citep{hoy39,bon44} serve as fundamental reference models. These theoretical approaches
provide a basis for comparison with the outcomes of numerical simulations \citep[e.g.,][]{ruf96} and are also useful when additional effects are introduced, such as
other gravitational potentials \citep[e.g.,][]{lin07,cio17,kaa19}, a rigid boundary around
the body \citep{thu16,pru24}, magnetic fields \citep[e.g.][]{igu02,tor12,san12,lee14},  
fluid instabilities \citep{bec18}, binaries \citep{ant19}, a low angular momentum \citep{pro03} or a turbulent
\citep{kru06,bur17,les23} medium.

Most studies on the BHL problem assume a large-scale, three-dimensional 
medium.
In realistic astrophysical systems, density or velocity gradients may be important 
\citep[e.g.,][]{liv86}.
In a spherical system with a radial density gradient, accretors in circular orbit 
feels a transverse density gradient; this occurs for instance in binary systems
\citep[e.g.,][]{mac15,xu19}. On the other hand,
there exist many situations where bodies are embedded in vertically-stratified discs.
For example, planets in protoplanetary discs, or black holes and stars within the accretion disc of AGNs.  In the case of
Bondi accretion, vertical statification is expected to reduce the accretion rate relative
to the homogeneous case \citep[e.g.,][]{dit21,zho24,che25}.
It should be noted, however, that embedded bodies, even if they are on circular orbits,  
accrete according to Bondi's prescription only when their masses lie
below the thermal mass \citep[e.g.][]{cho23}.
In particular, the resulting accretion flow onto gap-opening planets departs
significantly from the assumptions underlying Bondi accretion 
\citep[e.g.][]{ros20,li23}.

When planets, stars or black holes are on eccentric or inclined orbits, they typically move supersonically relative to the surrounding gas \citep[e.g.,][]{mut11,rei12,xia13}, placing them under the BHL accretion model.
An interesting scenario for BHL accretion within a disc arises
in the aftermath of the merger of two spiral galaxies, where intermediate-mass 
and supermassive black holes
migrate towards the galactic nucleus and can form an eccentric black-hole binary. 
This binary could be embedded within the nuclear 
gaseous disc of the merged galaxies \citep[see][for the case of equal-mass galaxies]{may07}.  
If the orbital eccentricities of these accretors are
$\gtrsim 2h$, where $h$ is the disc's aspect ratio, 
they will move supersonically with respect to the gas in the disc \citep[e.g.,][]{mut11}.
Black holes can move also supersonically if they are 
counterrotating with respect to the AGN accretion disc \citep{sec21,san25}.

A distinct case of BHL accretion in a disc emerges 
in a binary system formed by a high-mass Be star and a neutron star or
black hole. Due to its rapid rotation, the Be star may develop a decretion disc
\citep{han96,car11,mar25}. In eccentric binary systems, the neutron star (or
black hole) can accrete material from the disc when it approaches periastron
\citep[e.g.,][]{neg01,neg01b,oka02,oka13,li21}.

There are two key physical quantities of interest that one aims to estimate: 
the mass accretion rate onto the perturber and the drag force. The contribution
of the far-field wake to the drag force can be computed using linear theory.
Some authors have extended the linear-theory drag force formula from a homogeneous medium to a disc geometry by simply setting the maximum impact parameter to $\sim H$,
the scaleheight of the disc \citep[e.g.,][]{art93,xia13}.
\citet{mut11} compute the drag force on a body moving supersonically in a planar
2D medium in linear theory and apply it to study the migration and eccentricity
damping timescales of planets in eccentric
orbits \citep[see also][]{san19,ida20}. 
\citet{vic19} also use linear theory to compute the drag force and the 
wake generated by a body moving through a constant-density slab, imposing 
Dirichlet conditions on the upper and lower surfaces. 
One limitation of linear theory is the ambiguity in choosing the minimum impact
parameter, because the interaction becomes nonlinear at sufficiently small 
distances from the perturber. 

The ambiguity in the minimum impact radius can be resolved either by relying on 
numerical simulations \citep[e.g.,][]{ber13,suz24} or by modelling
the nonlinear regime directly.
\citet[][hereafter \citetalias{can13}]{can13} analytically evaluate the accretion mass rate 
and the drag force (including the contribution from the nonlinear inner part of the wake)
acting on a body embedded in a vertically stratified medium. In a thin layer, they find that
the total drag force is independent of the object's velocity.
For a thick layer, they determine the minimum and maximum impact parameters in the Coulomb logarithm. However, while \citetalias{can13} also employ a ballistic
approach and assume a zero thickness wake, as in the BHL model,
they additionally assume no mixing of wake material, in contrast with 
the BHL prescription, which assumes perfect mixing.
In the present work, using the classical BHL assumptions, we compute the tail density, accretion 
rate and drag force on a supersonic body moving in the midplane of 
a stratified medium.

The paper is organized as follows. Section \ref{sec:classical_overview} provides an overview of the BHL equations in the classical homogeneous medium within the BHL framework of line accretion.  In Section \ref{sec:BHL_stratified_medium}, 
we extend the BHL approach for a stratified medium, and obtain the continuity and momentum equations in the steady state. Section \ref{sec:thin_disk} focuses on the 
solution in the limit of a very thin layer, a case for which a simple analytical solution
can be derived. In Section \ref{sec:thick_disk}, we present the results obtained for layers
of arbitrary vertical thickness. A clarifying remark regarding the far-field wake’s contribution to the drag force is presented in Section \ref{sec:farfield}.
Finally, the conclusions are summarized in Section \ref{sec:conclusions}.

\section{The classical BHL problem: An overview}
\label{sec:classical_overview}
Since a homogeneous medium is a special case of a stratified medium with 
zero gradient, it is instructive to first consider the classical homogeneous case.
We treat the \citet{hoy39} and \citet{bon44} models in separate subsections, since they are not equivalent and may yield different 
mass accretion rates.

\subsection{The analysis of Hoyle \& Lyttleton (1939)}
\label{sec:HL_assumptions}
To study the accretion flow induced by a gravitating body of mass $M$ travelling 
through a cold homogeneous medium, \citet{hoy39} used a ballistic description of 
the gas and thus neglected the forces due to gas pressure.

Consider a coordinate system centred on the object, such that 
the gas approaches the body from large distances ($x\rightarrow -\infty$) with velocity 
$v_{0}\hat{\mathbf x}$, where $v_{0}>0$. Within this ballistic approximation,
the incoming particles move on hyperbolic orbits. The velocity components along 
the $x$-axis and in the perpendicular direction, for a 
streamline characterized by the impact parameter $\xi$ are given by
\begin{equation}
v_{x}=v_{0}\left(1+\frac{1}{\xi}|\sin \theta|\right),
\label{eq:vx_fs}
\end{equation}
and 
\begin{equation}
v_{y}=-\frac{v_{0}}{\xi} (1+\cos\theta),
\end{equation}
where $\theta$ is the polar angle measured from the downstream axis, such that
$\theta=0$ corresponds to the direction immediately behind the perturber.

These expressions describe the flow along a streamline until it intersects the $x$-axis.
A streamline with impact parameter $\xi$ intersects the $x$ axis at a distance
\begin{equation}
x_{\xi}=\frac{\xi^{2}}{2\xi_{0}}
\label{eq:x_xi}
\end{equation}
from the perturber. Here $\xi_{0}\equiv r_{\rm HL}/2=GM/v_{0}^{2}$, being 
$r_{\rm HL}$ the Hoyle-Lyttleton accretion radius $r_{\rm HL}\equiv 2GM/v_{0}^{2}$.
\citet{hoy39} argued that gas will collide in the tail behind the star and 
will lose kinetic energy because the $y$-component of the velocity thermalize.
They positioned that streamlines satisfying 
\begin{equation}
\frac{1}{2}v_{x}^{2} - \frac{GM}{x_{\xi}}<0
\end{equation}
will have insufficient energy to escape from the gravitational
potential of the star and therefore will be accreted onto the star. 
Since $v_{x}=v_{0}$ in the $x$-axis (Eq. \ref{eq:vx_fs}, with $\theta=0$), 
the above equation implies that all streamlines with impact parameter less than
$2\xi_{0}=r_{\rm HL}$ will be accreted. The mass accretion
rate is therefore given by 
\begin{equation}
\dot{M}_{\rm HL} 
=4\pi \rho_{0} v_{0}\xi_{0}^{2}= \frac{4\pi \rho_{0} G^{2}M^{2}}{v_{0}^{3}}.
\label{eq:dotM_HL}
\end{equation}
The above accretion rate is commonly refer to the Hoyle-Lyttleton accretion rate.

\citet{bon44} proposed a more refined description of the accretion flow. They argued that the Hoyle-Lyttleton accretion rate should be regarded
as an upper limit. In the following subsection,  
we present the equations governing the accretion tail in a homogeneous 3D medium.

\subsection{The BHL analysis for the tail in the case of a homogeneous 3D medium}
\label{sec:BHL_analysis}
\begin{figure}
  \includegraphics[width=1.0\columnwidth]{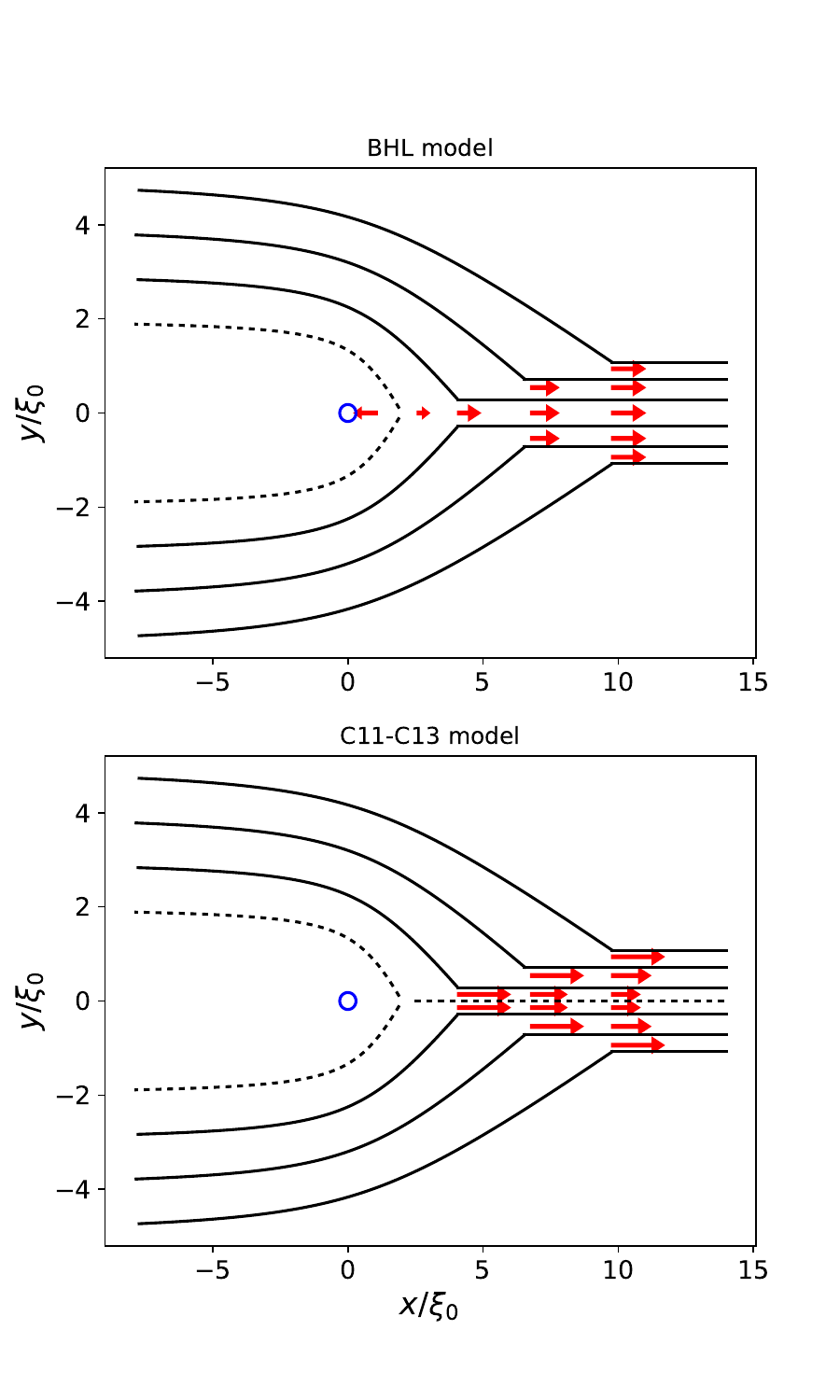} 
  \caption{Schematic of the flow (not to scale)
in the BHL approach (top panel) and in the C11-C13 model (bottom panel). 
The solid lines indicate the streamlines and the red arrows represent
the velocity in the tail. The dotted lines correspond to the streamline with impact
parameter $2\xi_{0}$. The tail is slower in the BHL model.
While the flow in the tail is assumed plane-parallel in both models,
it presents shear in the C11-C13 model. Streamlines with impact parameter
$\leq 2\xi_{0}$ are assumed to entry into the accretor, but they are not modelled 
in the C11- C13 model.}
  \label{fig:sketch}
\end{figure}

\citet{bon44} also neglected the pressure forces and assumed that,
when streamlines intersect, they form a 
high-density filament from the body downflow, i.e. at $x>0$. If energy is rapidly
irradiated, this accretion column or tail can be considered as infinitesimally
thin, with linear mass density $\mu(x)$ and velocity $v(x)$. Thus, $v(x)$ represents the
velocity of the material in the tail relative to the object. 
In fact, \citet{bon44} assumed that in the accretion column, all the material at a given $x$
moves with the same velocity $v(x)$ (see upper panel in Figure \ref{fig:sketch}).

To determine $\mu(x)$ and $v(x)$, \citet{bon44} derived the steady-state
continuity and momentum equations governing the structure of the tail. These equations
are given by
\begin{equation}
\frac{\partial (\mu v)}{\partial x} =2\pi \xi_{0} \rho_{0}v_{0},
\label{eq:continuity_3D}
\end{equation}
and
\begin{equation}
\frac{\partial (\mu v^{2})}{\partial x} = 2\pi \xi_{0} \rho_{0} v_{0}^{2} - \frac{GM\mu}{x^{2}},
\label{eq:momentum_3D}
\end{equation}
where $\rho_{0}$ is the density of the ambient medium
\citep[e.g.,][]{bon44,lyt72,edg04,mat15,rag22}.
The right-hand side of Equation (\ref{eq:continuity_3D}) and the first
term on the right-hand side of Equation (\ref{eq:momentum_3D}) represent the rates
of mass and momentum flux per unit length, being incorporated into the tail,
respectively. The second term in the right-hand side of Equation (\ref{eq:momentum_3D}) 
accounts for the gravitational force exerted by the object on the tail.
In this paper, we use the term `BHL model' to refer to the solutions of Equations 
(\ref{eq:continuity_3D})-(\ref{eq:momentum_3D}) in
a homogeneous medium, and to their extension to a stratified medium derived
in Section \ref{sec:BHL_stratified_medium}.

To solve the differential equations (\ref{eq:continuity_3D}) and (\ref{eq:momentum_3D}) 
for $\mu(x)$ and $v(x)$, \citet{bon44} imposed three boundary conditions:
(I) there is a stagnation point at $x=x_{0}$, i.e. $v=0$ at $x=x_{0}$,
(II) the velocity $v$ far downstream approach to $v_{0}$, and (III) $dv/dx \geq 0$
for all $x>0$.
\citet{lyt72} showed that there are infinitely many solutions that satisfy the boundary 
conditions imposed by \citet{bon44} and additionally found a class 
of `slow' solutions for which $v\rightarrow 0$ when $x\rightarrow \infty$. 
The different types of solutions including the effect of the pressure 
in the accretion column was studied by \citet{hor00}. 
\citet{edg04} showed that the condition (III) implies that $x_{0}\geq \xi_{0}$, i.e.
$\tilde{x}_{0}\equiv x_{0}/\xi_{0}\geq 1$.
\citet{mat15} removed the problem of the non-uniqueness of the solution 
by assuming three additional requirements: (IV) $d\mu/dx\geq 0$ for all $x>0$,
(V) $d^{2}v/dx^{2}\leq 0$ at all $x>0$ and (VI) $d^{2}\mu/dx^{2}\leq 0$.
They found that condition (IV) implies $1.325<\tilde{x}_{0}< 2$; condition (V) yields
$1.395<\tilde{x}_{0}< 1.768$; and condition (VI) constrains $\tilde{x}_{0}$
to  $1.617<\tilde{x}_{0}<1.627$.
Therefore, they conclude that the stagnation point in the 3D (axisymmetric) case
is located at $\tilde{x}_{0}\simeq 1.62$.

Recently, \citet{rag22} solve the
time-dependent equations using a finite-difference method until the solution reaches a
stationary state, obtaining $\tilde{x}_{0}\simeq 1.51$. Their solution resembles that of 
\citet{mat15}. \citet{rag22} also provide an analytical fit to
the solution.

The value of the stagnation point $\tilde{x}_{0}$ is physically relevant because it 
determines the mass accretion rate. From Eq. (\ref{eq:x_xi}), a streamline 
intersecting the $x$-axis at $x_{0}$
has an impact parameter $\xi_{\rm acc}=\sqrt{2\xi_{0}x_{0}}$. Since
$r_{\rm HL}=2\xi_{0}$, this reduces to $\xi_{\rm acc}=r_{\rm HL}$ only 
when $x_{0}=2\xi_{0}$; otherwise $\xi_{\rm acc}\neq r_{\rm HL}$. 
The mass accretion rate is given by
\begin{equation}
\dot{M} = \pi \rho_{0} v_{0}\xi_{\rm acc}^{2}  = \left(\frac{\tilde{x}_{0}}{2}\right)
\dot{M}_{\rm HL},
\end{equation}
where $\dot{M}_{\rm HL}$ is the nominal Hoyle-Lyttleton accretion rate
given in Eq. (\ref{eq:dotM_HL}). Consequently, if $\tilde{x}_{0}=1.62$, as derived by \citet{mat15}, $\dot{M}$ is reduced by a factor $0.81$ relative to $\dot{M}_{\rm HL}$.

In principle, numerical simulations can  adopt more realistic physical conditions and yield a more reliable estimate of the accretion rate. 
\citet{hor00} provide a compilation of
the values of $\tilde{x}_{0}$ reported in the literature \citep[see also][]{edg04}.

To evaluate the gravitational drag force experienced by $M$, we only
require $\mu(x)$. Although the free-streaming flow induces changes in the ambient density  
\citep[see Eqs. 7 and 8 in][hereafter \citetalias{can11}]{can11}, 
the density outside the thin tail does not contribute to the drag force on the body. 
This follows directly from Equation (\ref{eq:vx_fs}).
When a gas parcel reaches the symmetry axis ($\theta=0$),  its $x$-component
of the velocity remains $v_{0}$. 
Momentum conservation along the $x$-direction then ensures that the net $x$-force 
exerted by an entire streamline tube is zero\footnote{We caution that integrating the gravitational
force from gas in the free-streaming flow over a volume that contains incomplete 
streamline tubes (for example, a cylinder aligned with the $x$-axis) will 
yield a nonzero force.}.

\subsection{Underlying assumptions: BHL versus C11 and C13}
\label{sec:assumptions_BHL_C}
As mentioned in the previous subsection, the BHL model assumes that all material entering 
the tail mixes instantaneously and adopts a common velocity, with the associated dissipation
of energy.  Under this assumption, the velocity in the tail 
depends only on the distance along the symmetry axis and is uniform
accros any transverse cross section.
\citet{bon44} justified this approximation by arguing 
that small transverse velocity components 
generates circulation that efficiently mixes the gas within the cross section. 
For this assumption to hold, the associated transverse mixing timescale must
be shorter than the
characteristic time on which velocity gradients are produced by gravity.
This assumption is unlikely to be valid within the Hoyle-Lyttleton
accretion radius, 
where the velocity field inside the Mach cone exhibits a dependence on $\theta$ 
\citep[e.g.,][]{bis79}. However, it may provide a reasonable description of the flow farther downstream from the accretor, where the velocity field varies primarily along the axis \citep[e.g.,][]{ruf96}.

By contrast with the mixing assumption, \citetalias{can11} and \citetalias{can13}
assume that once the transverse velocity components, $v_y$ and $v_z$, are thermalized, the total energy along each streamline is conserved. As a result, the longitudinal component $v_x$ retains information about the fluid parcel’s original impact parameter, leading to a sheared velocity profile within the tail (see Figure \ref{fig:sketch}).
We will refer to the assumptions presented in references \citetalias{can11} and \citetalias{can13} as the C11–C13 model.

The calculation of the mass accretion rate
in \citetalias{can11} is identical to that used by \citet{hoy39} and described in
Subsection \ref{sec:HL_assumptions}. 
Consequently, in the \citetalias{can11} model, the accretion rate is equal to
$\dot{M}_{\rm HL}$. 

The assumptions underlying the BHL and C11-C13 are extreme limits, so
the true physical configuration is expected to be bracketed by these two limiting cases.
A conceptual caveat of \citet{hoy39} and C11-C13 models concerns
the motion of gas with impact parameters $\leq 2\xi_{0}$. As discussed
in C11, these models do not provide a clear description of
 how such gas initially moves away from the accretor before 
subsequently reverses direction and falls back toward the body.

\section{BHL equations in a Gaussian ambient medium}
\label{sec:BHL_stratified_medium}
We consider a gaseous medium stratified along the vertical direction $z$, with
a density profile given by  
\begin{equation}
\rho(z)= \rho_{0} \exp \left(-\frac{z^{2}}{2H^{2}}\right),
\label{eq:rho_z}
\end{equation}
where $\rho_{0}$ is the volume density at the midplane ($z=0$) and $H$
is the scaleheight. The surface density of the layer is $\Sigma=\sqrt{2\pi}\rho_{0}H$.
In the midplane, the accretor, which is represented as a point mass $M$, is
moving hypersonically at velocity $-v_{0}\hat{\bf{x}}$ relative to the ambient medium.
As in Section 
\ref{sec:classical_overview}, we take a coordinate system with the object at the centre, such that the gas upstream
at $x\rightarrow -\infty$ moves with velocity $v_{0}\hat{\mathbf x}$.
Note that, within the ballistic approximation, Equations (\ref{eq:vx_fs})-(\ref{eq:x_xi}) 
also remain valid in a vertically stratified medium.

Before outlining the BHL equations for $\mu(x)$ and $v(x)$ in
the stratified case, it is important to emphasize the limitations of the idealized
BHL assumptions when applied to a body embedded in a disc. 
In particular, the BHL framework assumes negligible gas pressure and 
vanishing angular momentum in the accretion flow. The ballistic approximation
is justified in the free-streaming region provided that the body moves supersonically relative
to the gas, a condition satisfied for eccentricities $e>2h$ \citep[e.g.,]
[]{pap02,mut11,ida20}. Within the downstream tail, pressure forces
become progressively less important as cooling becomes more efficient, i.e. as the sound speed in the tail tends to zero \citep{rag22};
this constitutes the most restrictive limitation of the BHL assumptions 
(see Section \ref{sec:farfield}).
On the other hand, background shear is negligible 
when the epicycle velocity exceeds the sound speed, so that the flow near the accretor is more nearly ballistic than shear dominated \citep[e.g.,][]{che22}. Indeed, 
in the supersonic regime ($e>2h$) and for a body with an orbital radius $r$, the response timescale of 
the dynamical-friction wake, $\sim
H/(e r\Omega)$, is shorter than the local shear timescale $\sim \Omega^{-1}$
\citep{pap02}, allowing the gas to react before differential rotation distorts the flow.

\subsection{Continuity and momentum equations}

To compute the mass flux feeding the tail, we first evaluate
the mass flux per unit impact parameter, which,
for the density distribution of Equation (\ref{eq:rho_z}) reads
\begin{align}
\frac{d\dot{m}}{d\xi} &= 4\rho_{0}v_{0}\xi \int_{0}^{\pi/2} \exp\left(-\frac{\xi^{2} \sin^{2}\theta}{2H^{2}}\right)\,d\theta 
\label{eq:dm_dxi}
\nonumber \\ & = 2\pi \rho_{0} v_{0}\xi \,I_{0}\left(\frac{\xi^{2}}{4H^{2}}\right)
\exp\left(-\frac{\xi^{2}}{4H^{2}}\right),
\end{align}
where $I_{0}(y)$ is the modified Bessel function of the first kind of order zero.
The amount of mass that is incorporated to the tail per unit distance, along $x>0$,
is given by
\begin{equation}
\Lambda (x) = \frac{d\dot{m}}{d\xi} \frac{d\xi}{dx}.
\end{equation}
Combining Eqs. (\ref{eq:x_xi}) and (\ref{eq:dm_dxi}), we obtain
\begin{equation}
\Lambda(x) = 2\pi \xi_{0} \rho_{0} v_{0} I_{0}\left(\frac{\xi_{0}x}{2H^{2}}\right)
\exp\left(-\frac{\xi_{0}x}{2H^{2}}\right).
\end{equation}

In the 1D approximation, the continuity equation is
\begin{equation}
\frac{\partial \mu}{\partial t} + \frac{\partial}{\partial x} (\mu v) = \Lambda.
\end{equation}
We recall that $\mu(x,t)$ is the linear mass density and $v(x,t)$ the velocity 
along the infinitesimally thin tail behind the accretor. 
On the other hand, the momentum equation is given by
\begin{equation}
\frac{\partial}{\partial t}(\mu v) +\frac{\partial}{\partial x}(\mu v^{2})=\Lambda v_{0}
-\frac{GM \mu}{x^{2}}.
\end{equation}

It is convenient to rewrite the above two equations in terms of dimensionless quantities
$\tilde{x}\equiv x/\xi_{0}$, $\tilde{\mu}=\mu/(\sqrt{2\pi}\Sigma\xi_{0})$ and
$\tilde{t}\equiv t v_{0}/\xi_{0}$. The resulting equations are
\begin{equation}
\frac{\partial \tilde{\mu}}{\partial \tilde{t}} + \frac{\partial}{\partial \tilde{x}} (\tilde{\mu} 
\tilde{v}) = \frac{1}{\tilde{H}} I_{0}\left(\frac{\tilde{x}}{2\tilde{H}^{2}}\right) 
\exp\left(-\frac{\tilde{x}}{2\tilde{H}^{2}}\right),
\label{eq:continuity_1}
\end{equation}
and
\begin{equation}
\frac{\partial}{\partial \tilde{t}}(\tilde{\mu} \tilde{v}) +\frac{\partial}{\partial \tilde{x}}
(\tilde{\mu} \tilde{v}^{2})=\frac{1}{\tilde{H}}I_{0}\left(\frac{\tilde{x}}{2\tilde{H}^{2}}\right) \exp\left(-\frac{\tilde{x}}{2\tilde{H}^{2}}\right)
-\frac{\tilde{\mu}}{\tilde{x}^{2}},
\label{eq:moment_1}
\end{equation}
where $\tilde{H}\equiv H/\xi_{0}$.

\subsection{Steady-state equations}
In the steady state, Equations (\ref{eq:continuity_1})  and (\ref{eq:moment_1})
are simplified to
\begin{equation}
 \frac{\partial}{\partial \tilde{x}} (\tilde{\mu} 
\tilde{v}) = \frac{1}{\tilde{H}} I_{0}\left(\frac{\tilde{x}}{2\tilde{H}^{2}}\right) 
\exp\left(-\frac{\tilde{x}}{2\tilde{H}^{2}}\right),
\label{eq:continuity_steady}
\end{equation}
and
\begin{equation}
\frac{\partial}{\partial \tilde{x}}
(\tilde{\mu} \tilde{v}^{2})=\frac{1}{\tilde{H}}I_{0}\left(\frac{\tilde{x}}{2\tilde{H}^{2}}\right) \exp\left(-\frac{\tilde{x}}{2\tilde{H}^{2}}\right)
-\frac{\tilde{\mu}}{\tilde{x}^{2}}.
\label{eq:moment_steady}
\end{equation}
Equations (\ref{eq:continuity_steady}) and (\ref{eq:moment_steady}) are the extensions
of Equations (\ref{eq:continuity_3D}) and (\ref{eq:momentum_3D}) to a Gaussian-stratified medium. We next combine Eqs. (\ref{eq:continuity_steady}) and (\ref{eq:moment_steady}) to obtain a differential equation for $\tilde{v}$ independent of $\tilde{\mu}$.

The above equation (\ref{eq:moment_steady}) can be written as
\begin{equation}
\tilde{v}\frac{d\tilde{\mu}\tilde{v}}{d\tilde{x}}+\tilde{\mu}\tilde{v}\frac{d\tilde{v}}{d\tilde{x}}
=\frac{1}{\tilde{H}}I_{0}(s) e^{-s}-\frac{\tilde{\mu}}{\tilde{x}^{2}}
\label{eq:moment_3}
\end{equation}
where  $s\equiv \tilde{x}/(2\tilde{H}^{2})$. Substituting $\partial(\tilde{\mu}\tilde{v})/d\tilde{x}$ from Eq. (\ref{eq:continuity_steady}) into Eq. (\ref{eq:moment_3}), we obtain
\begin{equation}
\tilde{\mu}\tilde{v}\frac{d\tilde{v}}{d\tilde{x}}
=\frac{1-\tilde{v}}{\tilde{H}}I_{0}(s) e^{-s}-\frac{\tilde{\mu}}{\tilde{x}^{2}}.
\label{eq:1plus2}
\end{equation}

In order to eliminate the dependence on $\tilde{\mu}$ in Equation (\ref{eq:1plus2}),
we  integrate Equation (\ref{eq:continuity_steady}), which yields
\begin{equation}
\tilde{\mu}\tilde{v}=2\tilde{H} (F(s)-F_{0}),
\label{eq:mu_v}
\end{equation}
with 
\begin{equation}
F(s)\equiv s e^{-s} [I_{0}(s)+I_{1}(s)],
\end{equation}
$F_{0}\equiv F(s_{0})$ and $s_{0}\equiv \tilde{x}_{0}/(2\tilde{H}^{2})$. From Eq.
(\ref{eq:mu_v}) we see that the velocity $\nu$ is zero at a distance $\tilde{x}_{0}$ 
and thus it corresponds to the stagnation point.

Finally, using Eq. (\ref{eq:mu_v}) into Eq. (\ref{eq:1plus2}), we obtain
\begin{equation}
\frac{d\tilde{v}}{d\tilde{x}}=\frac{(1-\tilde{v})I_{0}(s)e^{-s}}{2\tilde{H}^{2}[F(s)-F_{0}]}-\frac{1}
{\tilde{x}^{2}\tilde{v}}.
\label{eq:dv_dx_steady}
\end{equation}
This differential equation for $\tilde{v}$ does not explicitly depend on 
$\tilde{\mu}$, but does depend on $\tilde{x}_{0}$ through $F_{0}$. This equation
has a critical point at $s=s_{0}$ (or $\tilde{x}=\tilde{x}_{0}$). Before examining
the solution in the vicinity of this point, which we undertake in the next subsection,  
we demonstrate that, in the limit of a homogeneous medium, the equation simplifies to 
the standard BHL equation.

The case of a homogeneous medium corresponds to the limit of very large $\tilde{H}$.
In this limit, both $s$ and $s_{0}$ are very small so that we can expand $I_{0}(s)e^{-s}=
1-s+\mathcal{O}(s^{2})$, $I_{1}(s)e^{-s}=s/2 +\mathcal{O}(s^{2})$ and $F(s)=s+\mathcal{O}(s^{2})$ in Equations (\ref{eq:mu_v}) and (\ref{eq:dv_dx_steady}), to obtain
\begin{equation}
\tilde{\mu}\tilde{v} = \frac{1}{\tilde{H}} (\tilde{x}-\tilde{x}_{0}),
\label{eq:mu_v_3D}
\end{equation}
and
\begin{equation}
\frac{d\tilde{v}}{d\tilde{x}}=\frac{1-\tilde{v}}{\tilde{x}-\tilde{x}_{0}}-\frac{1}
{\tilde{x}^{2}\tilde{v}}.
\label{eq:dv_dx_3D}
\end{equation}
In the homogeneous 3D case, the conventional normalization for $\mu$ is $\mu_{\ast}=
\mu/(2\pi \rho_{0}\xi_{0}^{2})$. In these units, Equation (\ref{eq:mu_v_3D}) implies
\begin{equation}
\mu_{\ast} \tilde{v} = \tilde{x}-\tilde{x}_{0}.
\label{eq:hatmu_v}
\end{equation}
Equations (\ref{eq:dv_dx_3D}) and (\ref{eq:hatmu_v}) correspond to the Equations (8)
in \citet{bon44} for the homogeneous 3D case.

\subsection{Solution behaviour near the critical point}
It is possible to determine $\tilde{\mu}$ and $d\tilde{v}/d\tilde{x}$ at 
the critical point $\tilde{x}_{0}$ as follows.
In a neighbourhood of $\tilde{x}_{0}$, we have 
\begin{equation}
\tilde{v}\simeq (\tilde{x}-\tilde{x}_{0}) \frac{d\tilde{v}}{d\tilde{x}}\bigg|_{\tilde{x}_{0}}.
\label{eq:long}
\end{equation} 
Combining Eq. (\ref{eq:mu_v}) and (\ref{eq:long}), it holds that 
\begin{equation}
\tilde{\mu}\tilde{v}\simeq \tilde{\mu}_{0}\, (\tilde{x}-\tilde{x}_{0}) \frac{d\tilde{v}}{d\tilde{x}}\bigg|_{\tilde{x}_{0}}\simeq 2\tilde{H} (F(s)-F_{0}),
\label{eq:mu_der}
\end{equation}
where $\tilde{\mu}_{0}\equiv \tilde{\mu}(\tilde{x}_{0})$.
By substituting Eq. (\ref{eq:dv_dx_steady}) into Eq. (\ref{eq:mu_der}), it is simple to
show that 
\begin{equation}
\frac{\tilde{\mu}_{0}^{2}}{\tilde{x}_{0}^{2}}-\frac{\tilde{\mu}_{0}}{\tilde{H}} I_{0}(s_{0})e^{-s_{0}}=
-4\tilde{H}^{2}\lim_{\tilde{x}\rightarrow \tilde{x}_{0}}\frac{[F(s)-F_{0}]^{2}}{\tilde{x}-\tilde{x}_{0}}.
\label{eq:mu0_first}
\end{equation}
Since the limit in the RHS of Equation (\ref{eq:mu0_first}) is zero, the linear density 
at $\tilde{x}_{0}$ is
\begin{equation}
\tilde{\mu}_{0}= \frac{1}{\tilde{H}}\tilde{x}_{0}^{2} I_{0}(s_{0}) e^{-s_{0}}.
\label{eq:mu0_final}
\end{equation}
It is worth noting that in the above equations, $\tilde{x}_{0}$ implicitly depends on $\tilde{H}$.

Now we are going to evaluate $d\tilde{v}/d\tilde{x}$ at the critical point.
By expanding $F(s)\simeq F(s_{0}) + F'(s_{0}) (s-s_{0})$ (where the prime indicates derivative
respect to $s$), Equation (\ref{eq:mu_der}) can be recast as
\begin{equation}
\tilde{\mu}_{0} \frac{d\tilde{v}}{d\tilde{x}}\bigg|_{\tilde{x}_{0}} = \frac{1}{\tilde{H}}
\frac{dF}{ds}\bigg|_{s_{0}}.
\end{equation}
Noting that 
\begin{equation}
\frac{dF}{ds}\bigg|_{s_{0}} = e^{-s_{0}} I_{0}(s_{0}),
\end{equation}
and using Eq. (\ref{eq:mu0_final}), we get the simple relation 
\begin{equation}
\frac{d\tilde{v}}{d\tilde{x}}\bigg|_{\tilde{x}_{0}} = \frac{1}{\tilde{x}_{0}^{2}}.
\label{eq:dv_dx_x0}
\end{equation}

\section{Infinitely thin layer}
\label{sec:thin_disk}
We now specialize to the case of an infinitely thin layer (2D planar case): $\tilde{H}\ll 1$ and $\tilde{H}^{2}\ll
\tilde{x}$. The latter condition implies $s\rightarrow 0$. 
In this limit, Equations (\ref{eq:mu_v}), (\ref{eq:dv_dx_steady}) and (\ref{eq:mu0_final}) become
\begin{equation}
\tilde{\mu}\tilde{v}=\frac{2}{\sqrt{\pi}}(\sqrt{\tilde{x}}-\sqrt{\tilde{x}_{0}}),
\label{eq:mu_v_2D}
\end{equation}
\begin{equation}
\frac{d\tilde{v}}{d\tilde{x}}= \frac{ 1-\tilde{v} }
{2(\tilde{x}-\sqrt{\tilde{x}_{0}\tilde{x}})}
-\frac{1}{\tilde{x}^{2}\tilde{v}},
\label{eq:dv_dx_2D}
\end{equation}
and
\begin{equation}
\tilde{\mu}(\tilde{x}_{0}) = 
\frac{\tilde{x}_{0}^{3/2}}{\sqrt{\pi}}.
\label{eq:mu_at_x0_2D}
\end{equation}

Equations  (\ref{eq:mu_v_2D}), (\ref{eq:dv_dx_2D}), and (\ref{eq:mu_at_x0_2D})
were previously derived by \citet{sok90}, apart from some differences in factors due to 
the use of different variable normalization. However, as far as we know,
analytical solutions seem to have been overlooked.
After some algebraic manipulation, we find that the following expressions for $\tilde{\mu}$ and $\tilde{v}$ provide a simple analytical solution to Equations (\ref{eq:mu_v_2D})-(\ref{eq:mu_at_x0_2D})
\begin{equation}
\tilde{\mu}(\tilde{x}) =2\sqrt{\frac{\tilde{x}}{\pi}},
\label{eq:mu_2D}
\end{equation}
\begin{equation}
\tilde{v} (\tilde{x})= 1-\sqrt{\frac{2}{\tilde{x}}}.
\label{eq:v_exact_2D}
\end{equation}
This solution satisfies all the conditions (I to VI), outlined in Section 
\ref{sec:classical_overview}. Therefore, it is a physically aceptable solution for BHL tail,
within the simplyfing assumptions of the model.

Clearly, the critical point for this particular solution is $\tilde{x}_{0}=2$. 
While 2D planar simulations have been also conducted by many authors  \citep[e.g.,][]{anz87,mat87,taa88,mat91,blo13},
a large fraction of them focus on inhomogeneous media and 
they do not report the position of the stagnation point.

Once $x_{0}$ is known, the accretion radius can be derived from Eq. (\ref{eq:x_xi}), 
yielding $\xi_{\rm acc}=\sqrt{2\xi_{0}x_{0}}=2\xi_{0}$. 
It is remarkable that this value coincides with that derived by \citetalias{can11} and 
\citetalias{can13} using different assumptions. 
The accretion radius in the 2D planar
case appears slightly larger than in the 3D case, where it typically ranges from 
$1.74\xi_{0}$ \citep{rag22} to $1.8\xi_{0}$ \citep{mat15}.

The mass accretion rate in the 2D planar case is
\begin{equation}
\dot{M}=2\xi_{\rm acc}\Sigma v_{0}=\frac{4 GM\Sigma}{v_{0}}.
\end{equation}
This value of $\dot{M}$ matches the result obtained by \citetalias{can13}.

The accretion of mass produces a drag force, $F_{\rm acc}$, given by 
$\dot{M} v_{0}=4GM\Sigma$ on the accretor.
In addition, the tail exerts a gravitational force $F_{g}$ given by
\begin{equation}
F_{g} = G M \int_{x_{0}}^{x_{\rm max}} \frac{\mu}{x^{2}} dx.
\label{eq:grav_drag}
\end{equation}
Replacing Eq. (\ref{eq:mu_2D}) into the above equation and taking $x_{\rm max}=\infty$, 
we find
\begin{equation}
F_{g} = 2\sqrt{2} GM \Sigma \xi_{0}^{1/2} \int_{2\xi_{0}}^{\infty} x^{-3/2} dx = 
4 GM \Sigma.
\end{equation}
Therefore, the total drag force is $F_{T} = F_{\rm acc}+F_{g} = 8 GM\Sigma$.

While the value of $F_{\rm acc}$ is identical to the value derived in \citetalias{can13}, 
the magnitude of $F_{g}$ is slightly different. \citetalias{can13} find $F_{\rm g}=2 (\pi-2)GM$, which is smaller 
by a factor of $0.57$ compared to the value obtained here
under BHL assumptions. We attribute this discrepancy to the inherently dissipative
nature of the BHL (see Section \ref{sec:classical_overview}), which results in a denser and slower tail compared to
the C11-C13 model.

 \begin{figure}
  \includegraphics[width=\columnwidth]{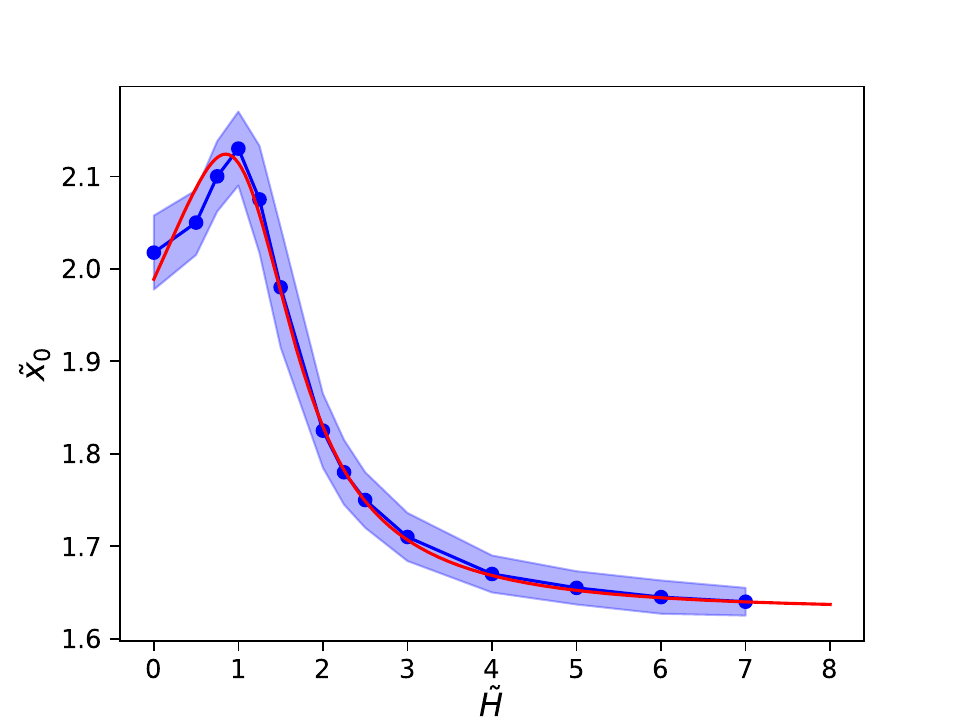}
  \caption{$\tilde{x}_{0}$ versus $\tilde{H}$. The blue band includes those values
of $\tilde{x}_{0}$ that satisfy conditions (I)-(VI). The dots indicate the central
value in that interval. The red line represents the analytical fit given in 
Eq. (\ref{eq:anlytical_x0_H}).}
  \label{fig:x0_vs_Htilde}
\end{figure}

\section{Disc with arbitrary $H$} 
\label{sec:thick_disk}
As with Equation (\ref{eq:dv_dx_3D}) derived by \citet{bon44} for a 3D 
homogeneous medium,
Equation (\ref{eq:dv_dx_steady}) also admits an infinite number of solutions
that satisfy the boundary conditions.
To find solutions to Eq. (\ref{eq:dv_dx_steady}) that satisfy conditions (I)-(VI),
we proceeded as follows. For a given pair of parameters $\tilde{H}$ and 
$\tilde{x}_{0}$, we integrated
Eq. (\ref{eq:dv_dx_steady}) inwards from $\tilde{x}=10$ to $\tilde{x}_{0}+0.02$,
applying an explicit fourth-order Runge-Kutta method. A shooting approach was 
employed, in which the
initial velocity at $\tilde{x}=10$ was adjusted to identify solutions (if any exist) that
fulfill conditions (I)-(VI) for the chosen values of $\tilde{H}$ and 
$\tilde{x}_{0}$. As a result, we identify a narrow region in the ($\tilde{H},\tilde{x}_{0}$) plane
within which physically admissible solutions at $\tilde{x}\geq \tilde{x}_{0}$ exist. 
To continue the integration inward, we smoothly pass through the critical point by taking
advantage of the fact that the functional form
\begin{equation}
\tilde{v}(\tilde{x}) = 1-\left(\frac{\tilde{x}_{0}}{\tilde{x}}\right)^{1/\tilde{x}_{0}}
\label{eq:v_function}
\end{equation}
satisfies Equation (\ref{eq:dv_dx_x0}) and, in addition, it matches the exact solution in the infinitely
thin layer (see Equation \ref{eq:v_exact_2D}). Moreover, \citet{rag22}
show that Equation (\ref{eq:v_function}) also provides a good fit to the solution in 
the homogeneous case. Thus,  the functional form in Eq. (\ref{eq:v_function}) is 
well-suited to describe the BHL flow close to the critical point. 
Specifically, we use Eq. (\ref{eq:v_function}) to compute $\tilde{v}$ at $\tilde{x}_{0}-0.02$, which serves as the initial velocity for the inward integration down to $\tilde{x}=0.19$,
carried out via the Runge-Kutta method. Finally, we verify that the resulting
solution $\tilde{v}(\tilde{x})$ satisfies conditions (I)-(VI) in the whole interval $[0.19,10]$
under consideration.

\begin{figure}
  \includegraphics[width=0.84\columnwidth]{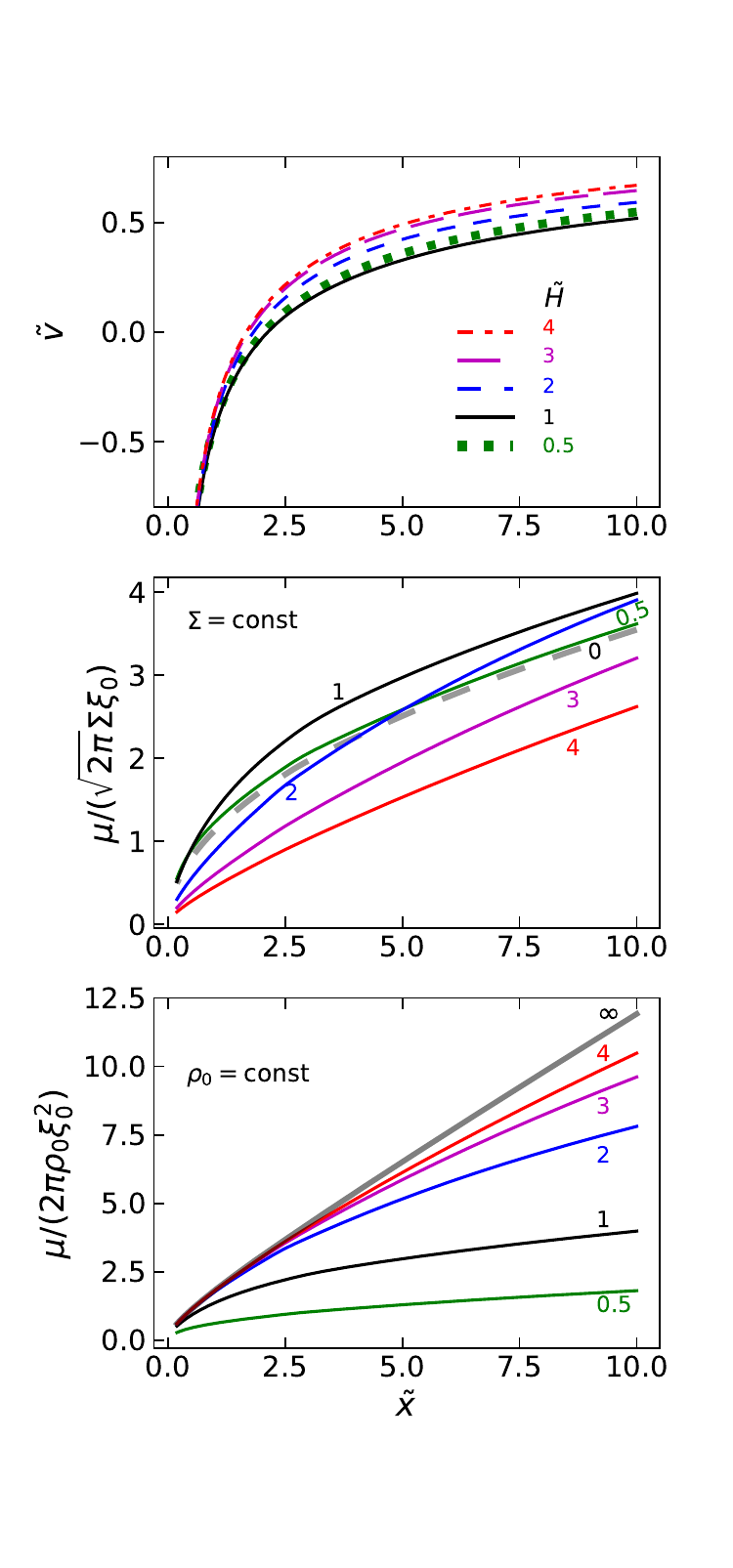}
  \caption{Dimensionless velocity $\tilde{v}$  (upper panel) and linear
density $\mu$, normalized to $\sqrt{2\pi}\Sigma\xi_{0}$ in the middle panel, and to $2\pi\rho_{0}\xi_{0}^{2}$ in the lower panel, shown for different values of $\tilde{H}$ indicated 
on each curve. For a given $\tilde{H}$, the value
of $\tilde{x}_{0}$ was chosen as the central point of the blue band in Figure \ref{fig:x0_vs_Htilde}.}
  \label{fig:v_and_mu}
\end{figure}

The narrow domain, delineating the viable
combinations of $\tilde{H}$ and $\tilde{x}_{0}$, is shown in Figure \ref{fig:x0_vs_Htilde}.
It is remarkable that $\tilde{x}_{0}$ has a maximum at $\tilde{H}\simeq 1$. 
At large $\tilde{H}$, $\tilde{x}_{0}$ tends to $1.62$, the value obtained for the homogeneous medium by \citet{mat15}, as expected.
The value of $\tilde{x}_{0}$ at the centre of the interval can be fitted by the function:
\begin{equation}
\tilde{x}_{0} (\tilde{H})= 1.63 + \frac{1}{3} \exp\left(-\frac{4\tilde{H}^{2}}{5}\right) + 
\frac{(\tilde{H}+0.2)^{8/5}}{\tilde{H}^{4}+3}.
\label{eq:anlytical_x0_H}
\end{equation}

\begin{figure}
  \includegraphics[width=0.84\columnwidth]{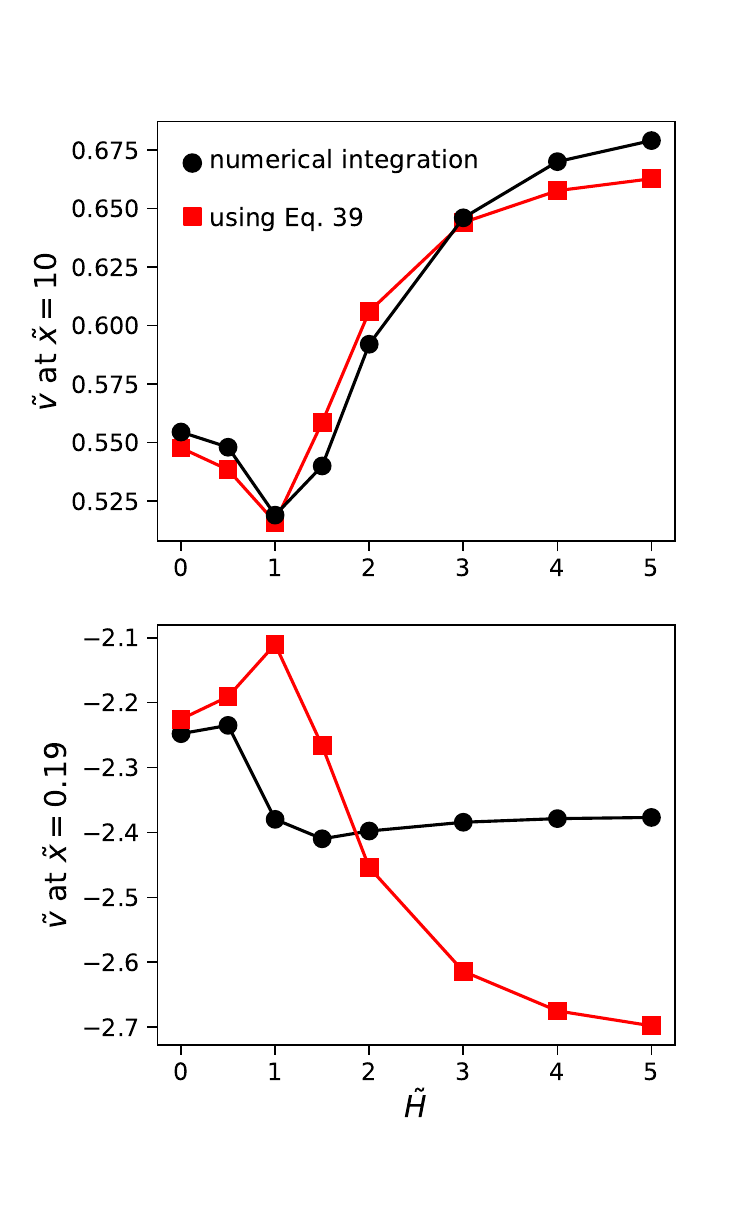} 
  \caption{$\tilde{v}$ at $\tilde{x}=10$ (top panel) and at $\tilde{x}=0.19$ (bottom
panel) as a function of $\tilde{H}$ in the solutions given in Figure \ref{fig:v_and_mu} (black circles), together with the predicted values using Equation (\ref{eq:v_function}) with the corresponding values
of $\tilde{x}_{0}$ (red squares).}
  \label{fig:comparison_v}
\end{figure}

Figure \ref{fig:v_and_mu} shows $\tilde{v}(\tilde{x})$ and $\tilde{\mu}(\tilde{x})$ for 
some values of $\tilde{H}$. For each $\tilde{H}$, the corresponding $\tilde{x}_{0}$ 
was chosen as the central value, indicated by the dots in 
Fig. \ref{fig:x0_vs_Htilde}\footnote{We note
tthat, given $\tilde{H}$ and the central value of $\tilde{x}_{0}$, there exists a narrow interval of values for $\tilde{v}$ at $\tilde{x}=10$ (the initial value in the shooting method) 
for which the solutions
satisfy conditions (I)-(VI). However, this interval is so narrow (with a width of approximately $0.002$) that the corresponding
solution curves appear indistinguishable in Figure \ref{fig:v_and_mu}. This ambiguity
in the value of $\tilde{v}$ at $\tilde{x}=10$ is less than the size of the symbols in the upper
panel of Figure \ref{fig:comparison_v}.}. This Figure shows
that, at least beyond $\tilde{x}=2$, $\tilde{v}(\tilde{x})$ increases 
as $\tilde{H}$ increases, provided $\tilde{H}>1$. At a fixed $\tilde{x}$,
the tail is slowest for $\tilde{H}=1$. This behaviour is illustrated
in the upper panel of Fig. \ref{fig:comparison_v}, where we see that
$\tilde{v}$ at $\tilde{x}=10$ is not a monotonic increasing function of $\tilde{H}$, but instead exhibits a minimum at $\tilde{H}=1$.
This behaviour is consistent with $\tilde{v}(\tilde{x})$ satisfying
Equation (\ref{eq:v_function}). Indeed, we have verified that  $\tilde{v}(\tilde{x})$ closely
adheres to this equation at $\tilde{x}>\tilde{x}_{0}$. The tail is slowest for $\tilde{H}=1$
because $\tilde{x}_{0}$ attains its maximum at that value of $\tilde{H}$ (see Figure \ref{fig:x0_vs_Htilde}).

In the inner region,
at $\tilde{x}<\tilde{x}_{0}$, the deviation of $\tilde{v}(\tilde{x})$ from Equation (\ref{eq:v_function}) is clearly visible in the lower panel of Fig. \ref{fig:comparison_v}. 
The values of $\tilde{v}$ at $\tilde{x}=0.19$ obtained from the numerical solution differ significantly from those predicted by Eq. (\ref{eq:v_function}), except for $\tilde{H}=0$ 
and $\tilde{H}=2$, where Equation (\ref{eq:v_function}) provides an excellent 
approximation within the interval
under consideration ($0.19\leq \tilde{x}\leq 10$). The accuracy of Eq. (\ref{eq:v_function}) 
for $\tilde{H}=0$ is expected, as it represents
an exact solution for $\tilde{H}=0$ and $\tilde{x}_{0}=2$, as demonstrated in Section
\ref{sec:thin_disk}.

While both the middle and lower panels of Figure \ref{fig:v_and_mu} display $\tilde{\mu}$ 
versus $\tilde{x}$, they do so using different units. In the middle panel, we normalize
to $\sqrt{2\pi}
\Sigma \xi_{0}$, which is useful when exploring situations where 
$\tilde{H}$ is varied while $\Sigma$ is held constant. In this sense, the middle panel shows
that at a given value of $\Sigma$, the BHL tail is densest for $\tilde{H}\simeq 1$, i.e.
$H\simeq \xi_{0}$. Note that the curves of $\tilde{\mu}$ versus $\tilde{x}$ in the middle
panel can intersect.

\begin{figure}
  \includegraphics[width=0.84\columnwidth]{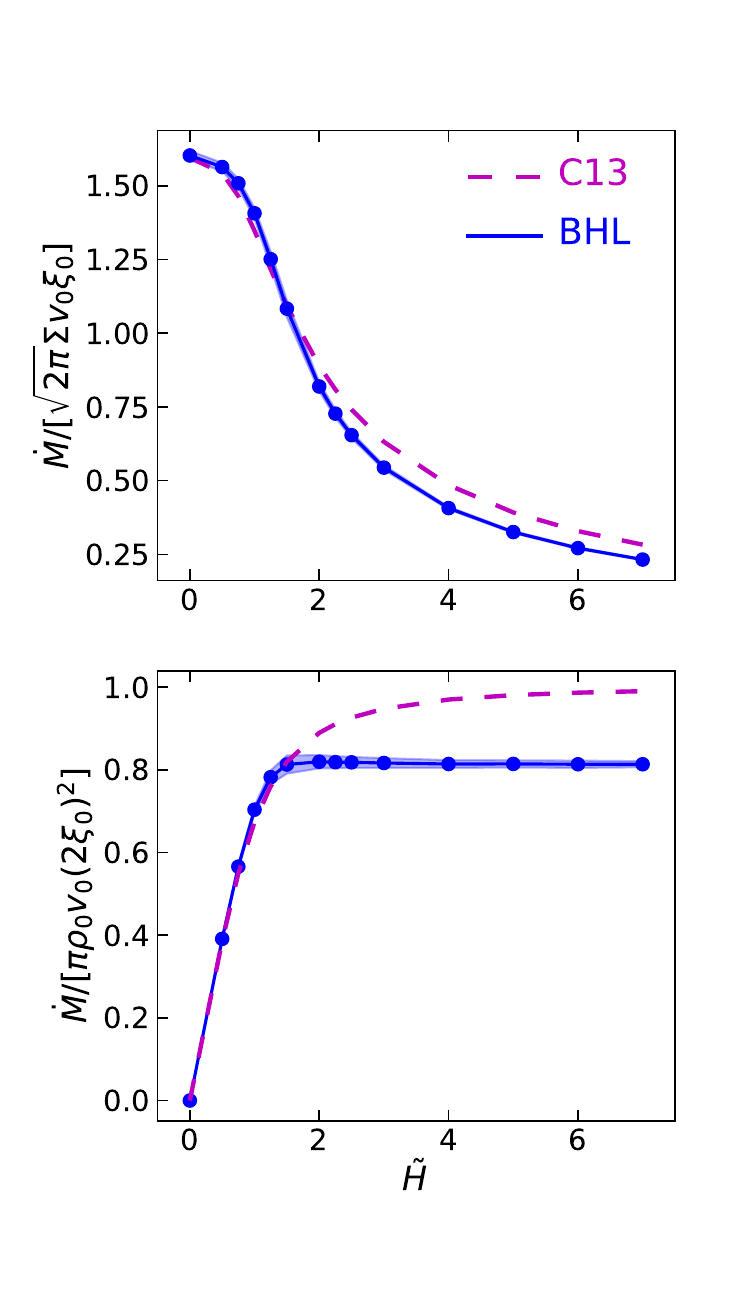} 
  \caption{Mass accretion rate as a function of $\tilde{H}$. In the top panel, the accretion
rate is normalized to $\sqrt{2\pi} \Sigma v_{0} \xi_{0}$, whereas it is normalized to 
$\pi \rho_{0}v_{0}(2\xi_{0})^{2}$ in the bottom panel. The solid curves correspond to
the BHL model whereas the dashed lines for C13 model.}
  \label{fig:accretion_rate}
\end{figure}

The lower panel is useful for examining how the linear density $\mu$ depends on the layer
thickness when the volume density in the midplane remains fixed. 
As $\tilde{H}$ increases, the curves asymptotically approach to the solution 
corresponding to the classical homogeneous medium (ie. case $\tilde{H}\rightarrow \infty$).
It is apparent that the tail formed by an accretor in a thin layer is less dense than in
a thick layer. 
While $\tilde{\mu}$ changes by nearly a factor of $2$ when $\tilde{H}$ increases
from $1$ to $2$, it varies by only a factor of $1.18$ when $\tilde{H}$ increases from $2$ to
$3$.

The mass accretion rate onto the body can be evaluated as the mass flow rate
with impact parameter less than $\xi_{\rm acc}\equiv \sqrt{2\xi_{0} x_{0}}$. More specifically,
\begin{equation}
\dot{M}=\int_{0}^{\xi_{\rm acc}} \frac{d\dot{m}}{d\xi}\, d\xi
=2\pi \tilde{x}_{0} \rho_{0} v_{0} \xi_{0}^{2} e^{-s_{0}} \left[I_{0}(s_{0})+I_{1}(s_{0})\right],
\end{equation}
with $d\dot{m}/d\xi$ given in Equation (\ref{eq:dm_dxi}).
Figure \ref{fig:accretion_rate}
 shows $\dot{M}$ as a function of $\tilde{H}$ using two normalizations. If $\Sigma$
is held constant and $\tilde{H}$ is varied, the maximum accretion rate occurs 
for $\tilde{H}=0$. At large $\tilde{H}$,
$\dot{M}$ decreases because the volume density at the midplane, where the accretor
is embedded, decreases as $\rho_{0}=\Sigma/(\sqrt{2\pi}H)\propto H^{-1}$.
An analytical fit of $\dot{M}$ versus $\tilde{H}$ yields the following result
\begin{equation}
\frac{\dot{M}}{\sqrt{2\pi} \Sigma v_{0} \xi_{0}} = \frac{2.9}{0.63\tilde{H}^{3/2}+1.9}+
0.3 (\tilde{H}+0.2) e^{-\tilde{H}^{2}/2}.
\end{equation}
Conversely, when $\rho_{0}$ is fixed and only $\tilde{H}$ is varied,
$\dot{M}$ remains roughly constant beyond $\tilde{H}\simeq 2$ (see lower panel
in Fig. \ref{fig:accretion_rate}).
Below this value, $\dot{M}$ decreases with decreasing $\tilde{H}$ due to the reduced
mass of the layer ($\Sigma \propto H$). In Figure \ref{fig:accretion_rate}, we also
show the accretion rate predicted by the \citetalias{can13} model. It agrees with the
BHL value for $\tilde{H}\lesssim 2$, but exceeds it by about $25\%$ at large 
$\tilde{H}$ values. The lower panel reveals that, at large $\tilde{H}$, the dimensionless
accretion rates converge to $0.81$ for the BHL model and to unity for the C13 model, as described in Sections
\ref{sec:BHL_analysis} and \ref{sec:assumptions_BHL_C}.

\begin{figure*}
  \includegraphics[width=1.87\columnwidth]{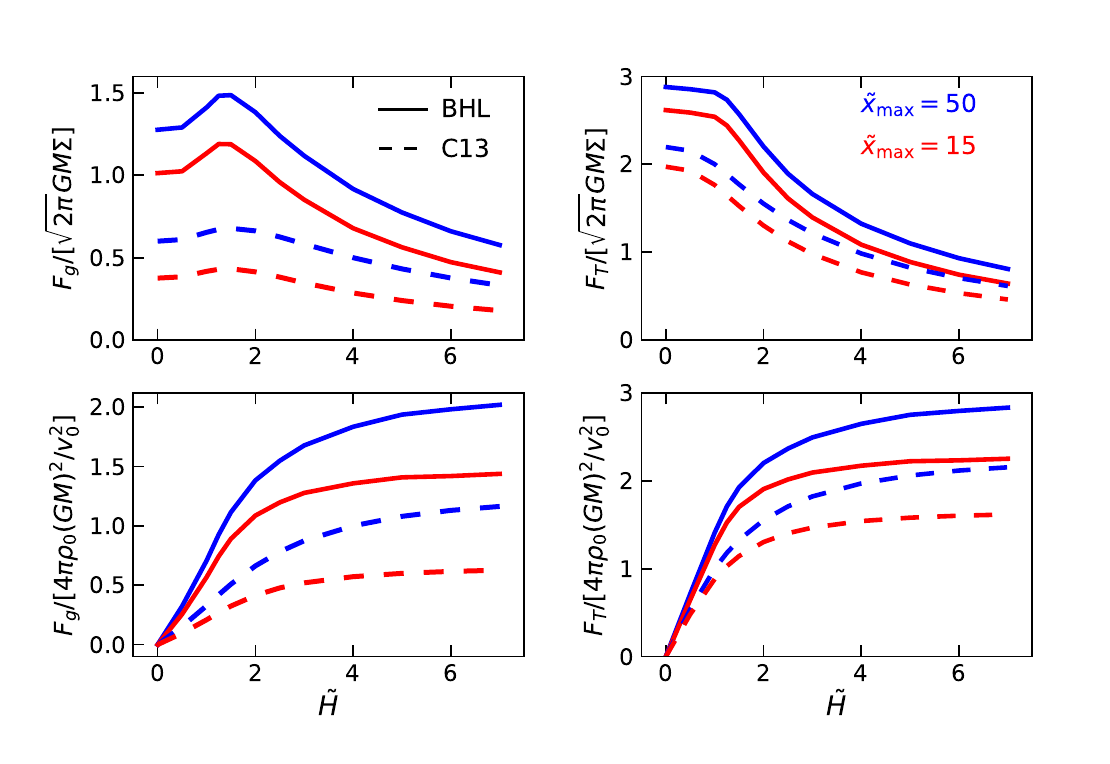} 
  \caption{Gravitational drag $F_{g}$ (left column) and the total drag $F_{T}$ (gravitational
plus accretion contributions; right column) for two different
values of $\tilde{x}_{\rm max}$, as a function of $\tilde{H}$.}
  \label{fig:drag_force_H}
\end{figure*}

Once $\mu(x)$ is known, we can compute the gravitational drag $F_{g}$ using 
Equation (\ref{eq:grav_drag}). The total drag $F_{T}$ will be $F_{g}+F_{\rm acc}$, with
$F_{\rm acc}=\dot{M}v_{0}$. Figure \ref{fig:drag_force_H} shows $F_{g}$ and $F_{T}$
as a function of $\tilde{H}$, for $x_{\max}=15\xi_{0}$ and  $50\xi_{0}$. 
The left upper panel of Fig. \ref{fig:drag_force_H} indicates that, if $\xi_{0}$
and $\Sigma$ are both fixed, then $F_{g}$ has a maximum at $\sim 1.5 \tilde{H}$. For
values larger than $1.5\tilde{H}$, $F_{g}$ decreases monotonically.  
The lower panels show $F_{g}$ and $F_{T}$ normalized to 
$4\pi \rho_{0} (GM)^{2}/v_{0}^{2}$. As expected, if $\rho_{0}$ is kept constant and
$\tilde{H}$ increases, $F_{g}$ increases as we have more material in the surrounding
medium to be incorporated into the tail. 

The shape of $F_{g}$ as a functin of $\tilde{H}$ resembles that found in \citetalias{can13}
(see Figure \ref{fig:drag_force_H}).
However, as in the $\tilde{H}=0$ case, $F_{g}$ is nearly twice as large as in
the \citetalias{can13} model. Nevertheless, because $F_{\rm acc}$ is similar in both
approaches, the total drag from the BHL tail is 
about $30\%$ larger than in \citetalias{can13}.

For accretors embedded in accretion discs on (quasi-)elliptical orbits, 
the timescales for radial migration and eccentricity damping can be estimated using a local approximation, in which
$\vecF_{T}$ is evaluated pointwise along the orbit. This analysis is presented in the Appendix, where
we also compare the results with previous studies that address the high-eccentricity regime.

\begin{figure}
  \includegraphics[width=0.9\columnwidth]{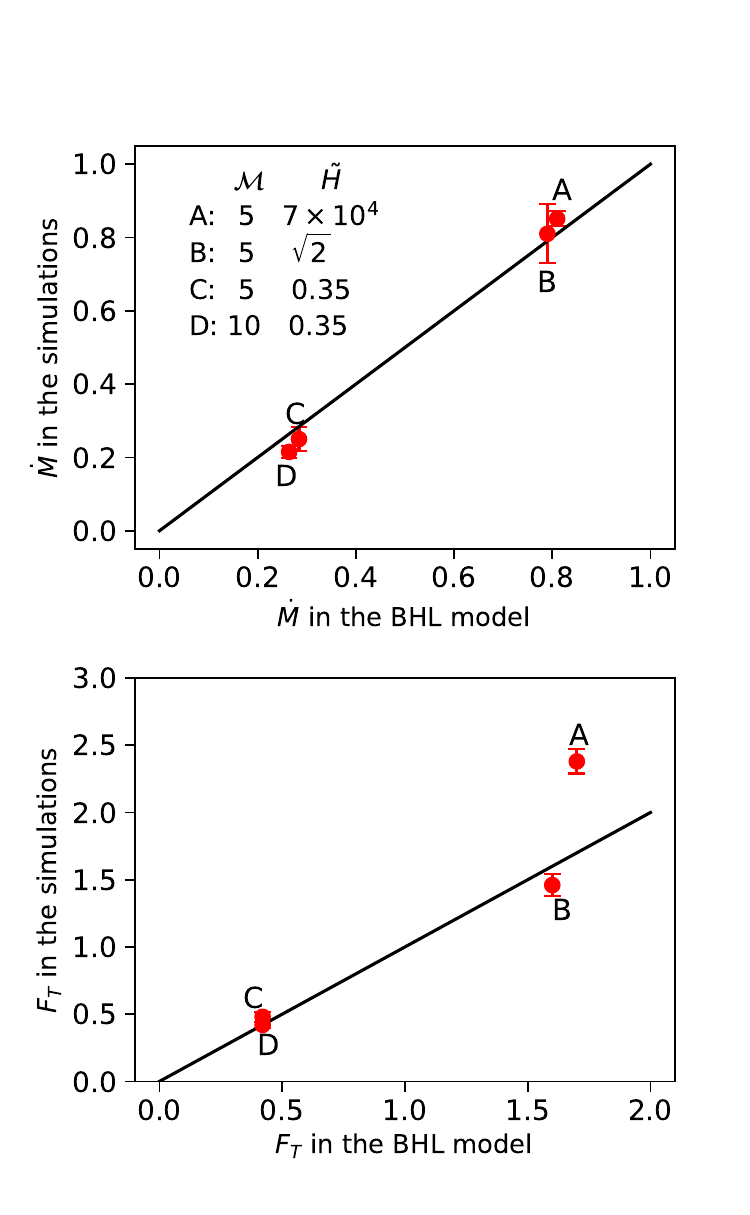} 
  \caption{One-to-one comparison between the mass accretion rate $\dot{M}$ (upper
panel) and total drag force $F_{T}$ (lower panel)
measured in the simulations
of \citetalias{can13} (vertical axis) and the corresponding predictions of the BHL model
(horizontal axis). The solid lines represent the identity relation. $\dot{M}$ is given in
units of $\pi \rho_{0} v_{0} (2\xi_{0})^{2}$, whereas the forces are expressed in
units of $4\pi \rho_{0}(GM)^{2}/v_{0}^{2}$. In the upper panel, the symbols have been
slightly offset horizontally to avoid overlap.}
  \label{fig:drag_sims_BHL}
\end{figure}

To conclude this section, we compare the BHL predictions with the simulations of \citetalias{can13}.
These simulations consist of four 3D simulations with Mach number $\mathcal{M}
\equiv v_{0}/c_{s}\geq 5$ and different values of $\tilde{H}$.  The computational 
domain extends from $-7.5\xi_{0}$ to $7.5\xi_{0}$ in the direction
of motion, so that $\tilde{x}_{\rm max}=7.5$.
Figure \ref{fig:drag_sims_BHL} compares the simulated mass accretion rates
$\dot{M}$ and the total drag forces $F_{T}$ with the corresponding BHL predictions. 
The BHL model reproduces the mass accretion rate accurately.
Its estimates of $F_{T}$ in models B, C, and D,
all of which satisfy $\tilde{H}\leq \sqrt{2}\xi_{0}$, also agree reasonably well with the simulations. In contrast,  for model A (nearly homogeneous case), 
the BHL model underestimates the drag force by a factor of about $1.4$.
\footnote{The C13-model also underestimates $F_{T}$. Although \citetalias{can13} 
report good agreement for model A, their result relies on a finite-box correction 
applied to the impact parameter, whereas it should instead be applied to the wake 
extent, as done here.}

\section{Far-field gravitational drag}
\label{sec:farfield}

The most significant limitation of BHL framework is that it neglects pressure effects.
In the previous sections, we have compared two models that share this
approximation but differ in their assumptions about the degree of mixing in the tail.
This distinction alone leads to a factor-of-two difference in the resulting $F_{g}$. 
On basic grounds, one expects the pressure term to modify the tail structure
beyond a few $\xi_{0}$.  
On the other hand, linear theory provides a good description of the far-field wake.
Comparing $\mu$ or $F_{g}$ predicted by the BHL model with that from linear theory therefore
provides a direct test of how well BHL captures the far-field wake.

In linear theory, the (normalized) steady-state density induced by a body moving hypersonically ($\mathcal{M}\gg 1$) in a homogeneous 3D medium is given by
\begin{equation}
\frac{\rho}{\rho_{0}}=1+\frac{GM}{c_{s}^{2}}\frac{2}{(x^{2}-\mathcal{M}^{2}R^{2})^{1/2}}
\label{eq:rho_lin}
\end{equation}
at the Mach cone ($x>\mathcal{M}R$), and $1$ outside of it
\citep[e.g.,][]{dok64,rep80,jus90,ost99}. The last expression is properly valid
if $x^{2}-\mathcal{M}^{2}R^{2}\gg (GM/c_{s}^{2})^{2}$, implying 
$x\gg \mathcal{M}\xi_{0}$. From Equation (\ref{eq:rho_lin}), we obtain that
the column density at $x\gg \mathcal{M}\xi_{0}$ that contributes to the gravitatioal drag is
\begin{equation}
\mu(x) = 2\pi \int_{0}^{x/\mathcal{M}} (\rho-\rho_{0}) R\,dR= 4\pi \xi_{0} \rho_{0} x. 
\end{equation}
Therefore, linear theory predicts
\begin{equation}
\frac{\mu}{2\pi \rho_{0}\xi_{0}^{2}}=2\tilde{x},
\end{equation}
whereas the BHL framework yields
\begin{equation}
\frac{\mu}{2\pi \rho_{0}\xi_{0}^{2}}\simeq \frac{\tilde{x}-\tilde{x}_{0}}{1-(\tilde{x}_{0}/x)^{1/\tilde{x}_{0}}}\simeq \tilde{x}
\end{equation}
in the far field
\citep[see bottom panel of Fig. \ref{fig:v_and_mu} and][]{rag22}. Thus, for 
wakes much longer than $\gg \mathcal{M}\xi_{0}$, the BHL model underestimates
the far-field gravitational drag in the homogeneous 3D case.

Now consider the limiting case of a 2D planar medium ($H\rightarrow 0$).
From Equation (41) in \cite{mut11},  linear theory predicts that 
the gravitational drag force exerted by the portion of the wake between $x=\mathcal{M}\xi_{0}$ and $x=\infty$ 
is $\delta F_{g}\simeq \pi GM\Sigma/\mathcal{M}$. In contrast, within the BHL framework, it is 
\begin{equation}
\delta F_{g} = 2\sqrt{2} GM \Sigma \xi_{0}^{1/2} \int_{\mathcal{M}\xi_{0}}^{\infty} x^{-3/2} dx = 
4\sqrt{\frac{2}{\mathcal{M}}}GM\Sigma
\end{equation}
(see Section \ref{sec:thin_disk}).
Hence, for an infinite-length wake in the 2D planar case and for $\mathcal{M}=5$, 
the BHL model overestimates $F_{g}$ by nearly a factor of $2$. 

\section{Concluding remarks}
\label{sec:conclusions}
In this paper, we have studied the flow structure (linear density, velocity and stagnation
point position $x_{0}$), the mass accretion rate and the drag
force on an accretor that moves in a straight line in the midplane of a vertically
stratified environment. As in the classical BHL problem involving a homogeneous
medium, we assume that
the tail is sufficiently thin to be described by a one-dimensional model.
Similarly to the homogeneous case, there exist infinite solutions that satisfy the boundary conditions.  

In the limit of an infinitely thin layer, we have found a simple analytical solution that
satisfies conditions (I)-(VI). 
This analytical solution may serve as a benchmark for future studies and as a valuable reference for pedagogical purposes. 

For finite thickness, the solution was obtained
by performing the integration inward and employed a shooting method to identify 
solutions that satisfy conditions (I)-(VI) described in Section \ref{sec:classical_overview}. This approach allowed 
us to determine a narrow region in the parameter space $(\tilde{H}, \tilde{x}_{0})$ 
where these conditions are met. Within this region, the central value of 
$\tilde{x}_{0}$ as a function of $\xi_{0}$ and $H$ has been provided.
The central value of $\tilde{x}_{0}$ reaches a maximum at $\tilde{H}=1$. 
For such a value of $\tilde{H}$, and at fixed $\Sigma$, the tail is both densest and slowest. 
In the limit of an infinitely thick layer,
we recover the BHL solution found by \citet{mat15} for a homogeneous medium.   

We have also computed the accretion rate $\dot{M}$ and derived a fitting formula.
If the surface density of the layer is kept constant, the maximum accretion occurs
in the case of an infinitely thin layer. If the midplane density $\rho_{0}$ is held constant
and $\tilde{H}$ is varied, $\dot{M}$ remains approximately constant for $\tilde{H}\gtrsim 2$. We have found that $\dot{M}$ is slightly smaller in
the BHL approach than in the C11-C13 model when $H>2\xi_{0}$. The values predicted
by the BHL model are in good agreement with those obtained in the simulations of 
\citetalias{can13}.

We have further determined the drag force due to accretion and the gravitational drag
force by the part of the tail that is not being accreted. By comparing with the simulations
of \citetalias{can13}, we have found that the BHL approach underestimates the total drag
force for thick layers. For thin layers ($H\leq 1.4\xi_{0}$), 
the BHL framework reproduces the values of $F_{T}$ contributed by the 
nonlinear part of the tail, 
although it is expected to overestimate the total drag when the tail becomes very long.

In addition to neglecting the pressure effects implied by the BHL assumptions, our calculations are based on further simplifying approximations.
First, we neglect the curvature
effects that arise when perturbers move along Keplerian orbits within discs around
a massive central object. We also ignore radial density gradients in the disc. 
Moreover, sufficiently massive perturbers can 
modify both the surface density and the velocity field in the disc, leading to gaps and altered
flow patterns. When gaps form, the modified profiles must be used instead of the unperturbed ones, which makes the problem significantly more difficult to handle.

\section*{Acknowledgements}
The author thanks the referee for a very detailed and constructive report, and Jorge Cant\'o for carefully reading the manuscript.

\section*{Data Availability}
This study does not involve the use or production of original data.

\appendix
\section{Characteristic timescales for bodies in elliptical orbits}
 
We consider an accretor of mass $M$ embedded in the midplane of a protoplanetary disc around a solar-mass star, 
on an orbit with semi-major axis $a$ and eccentricity $e$.
Following \citet{mut11} \citep[see also][]{san20}, the timescales for radial migration $t_{a}$, and eccentricity damping $t_{e}$, defined as
\begin{equation}
t_{a}\equiv \frac{a}{\left<da/dt\right>},
\end{equation}
and
\begin{equation}
t_{e}\equiv \frac{e}{\left<de/dt\right>}
\end{equation}
can be evaluated in the local approximation. In these expressions, the angle brackets $\left<.\right>$ denote an average over one orbit. Note that,
in our convention, the timescale is negative if the associated orbital element decreases with time.

\begin{figure*}
  \includegraphics[width=1.87\columnwidth]{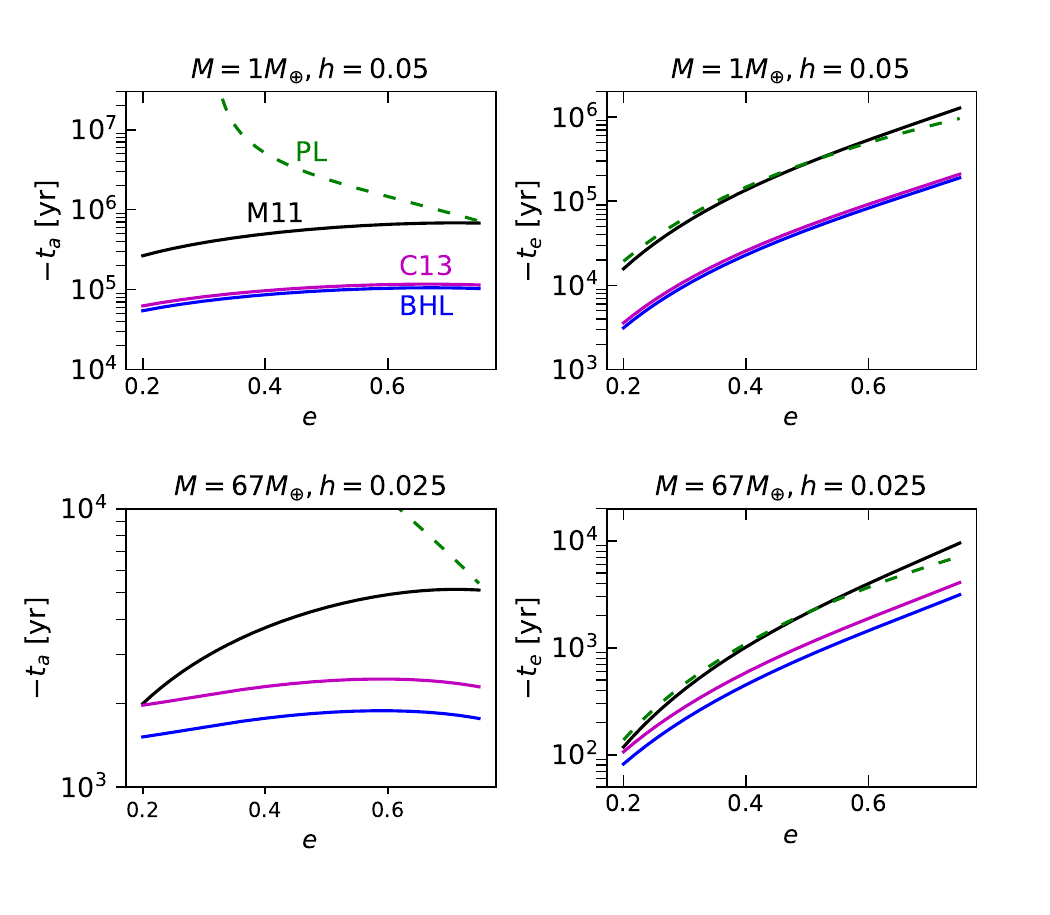} 
  \caption{Radial migration timescale $t_{a}$ (left column) and eccentricity damping timescale $t_{e}$ (right column) as functions of eccentricity. The upper panels correspond to  $M=1M_{\odot}$ and $h=0.05$, while the lower panels show results for $M=67M_{\oplus}$ and $h=0.025$. In the upper left panel, curves are labeled according to the underlying prescription: PL denotes the model of \citet{pap00}, and M11 that of \citet{mut11}. For a given set of parameters, the shortest timescales are obtained with the BHL model.}
  \label{fig:ta_te}
\end{figure*}

To facilitate comparison with previous studies, we consider a disc with surface density $\Sigma\propto r^{-3/2}$, a constant aspect ratio $h$, and a disc mass of $2M_{J}$ (where $M_{J}$ is the Jupiter mass) contained within $5$ au. Figure \ref{fig:ta_te} shows $t_{a}$ and 
$t_{e}$ as a function of $e$ for a body with $a=1$ au, and two different values of $M$ and $h$. 
To compute the BHL total drag force, we must specify the cut-off parameter $\tilde{x}_{\rm max}$.
For the case $M=1M_{\oplus}$ and $h=0.05$, we adopt $\tilde{x}_{\rm max}=200$, whereas for $M=67M_{\oplus}$ and $h=0.025$, we take
$\tilde{x}_{\rm max}=50$.
We have also plotted results using the dynamical friction force derived by \citet{mut11} (their Equation 41)
based on linear theory in a 2D disc, as well as the corresponding timescales from \citet{pap00}, computed using their Equations (31) and (32). We note
that \citet{pap00} do not provide $t_{a}$ directly, but instead give $t_{m}$, the timescale for changes in angular momentum. 
Nevertheless, $t_{a}$
can be readily obtained using
\begin{equation}
2t_{a} = \left(t_{m}^{-1} +\frac{e^{2}}{1-e^{2}} t_{e}^{-1}\right)^{-1}.
\end{equation}

For $M=1M_{\oplus}$ and $h=0.05$, $\tilde{H}$ is very large ($\tilde{H}\gtrsim 150$ for $e=0.2$ and 
$\tilde{H}\gtrsim 2000$ for $e=0.75$, at all points along the orbit).
Since the force increases logarithmically on $\tilde{H}$ in both C13 and BHL, the two approaches predict very similar timescales. 
For $M=67M_{\oplus}$ and $h=0.025$, the BHL model yields slightly shorter timescales than C13. We warn, however, that for this combination of parameters the local approximation is unlikely to be a good assumption \citep{san19}.

We find that C13 and BHL predict timescales that are significantly shorter than those obtained by \citet{pap00} and \citet{mut11} because these latter studies account only for the disc response at distances larger than $\sim H$, neglecting the contribution of the wake within the disc
thickness \citep[see also][]{san18}. As a result, they effectively provide upper limits to the migration and damping timescales.

\begin{thebibliography}{99}
\bibitem[Antoni et al.(2019)]{ant19}
Antoni, A., MacLeod, M., Ramirez-Ruiz, E. 2019, \apj, 884, 22
\bibitem[Anzer et al.(1987)]{anz87}
Anzer, U., Borner, G., Monaghan, J. J. 1987, \aap, 176, 235
\bibitem[Artymowicz et al.(1993)]{art93}
Artymowicz, P., Lin, D. N. C., Wampler, E. J. 1993, \apj, 409, 592
\bibitem[Beckmann et al.(2018)]{bec18}
Beckmann, R. S., Slyz, A., Devriendt, J. 2018, \mnras, 478, 995
\bibitem[Bernal \& S\'anchez-Salcedo(2013)]{ber13}
Bernal, C. G., S\'anchez-Salcedo, F. J. 2013, \apj, 775, 72
\bibitem[Bisnovatyi-Kogan et al.(1979)]{bis79}
Bisnovatyi-Kogan, G. S., Kazhdan, Ya. M., Klypin, A. A., Lutskii, A. E., Shakura, N. I. 
1979, Soviet Astronomy, 23, 201
\bibitem[Blondin(2013)]{blo13}
Blondin, J. M. 2013, \apj, 767, 135
\bibitem[Bondi(1952)]{bon52}
Bondi, H. 1952, \mnras, 112, 195
\bibitem[Bondi \& Hoyle(1944)]{bon44}
Bondi, H., Hoyle, F. 1944, \mnras,  104, 273
\bibitem[Burleigh et al.(2017)]{bur17}
Burleigh, K. J., McKee, C. F., Cunningham, A. J. 2017, \mnras, 468, 717
\bibitem[Cant\'o et al.(2011)]{can11}
Cant\'o, J., Raga, A. C.,, Esquivel, A., S\'anchez-Salcedo, F. J. 2011, \mnras, 418, 1238 (C11)
\bibitem[Cant\'o et al.(2013)]{can13}
Cant\'o, J., Esquivel, A., S\'anchez-Salcedo, F. J., Raga, A. C. 2013, \mnras, 762, 21 (C13)
\bibitem[Carciofi(2011)]{car11}
Carciofi, A. C. 2011, in Neiner C., Wade G., Meynet G., Peters G.eds, Proc.
IAU Symp. 272, Active OB Stars: Structure, Evolution, Mass Loss, and
Critical Limits. Cambridge University Press, Cambridge, p. 325
\bibitem[Chen et al.(2022)]{che22}
Chen, Y.-X., Bailey, A., Stone, J., Zhu, Z. 2022, \apjl, 939, L23
\bibitem[Chen et al.(2025)]{che25}
Chen, Y.-X., Jiang, Y.-F., Goodman, J. 2025, \apj, 987, 188
\bibitem[Choksi et al.(2023)]{cho23}
Choksi, N., Chiang, E., Fung, J., Zhu, Z. 2023, \mnras, 525, 2806
\bibitem[Ciotti \& Pellegrini(2017)]{cio17}
Ciotti, L., Pellegrini, S. 2017, \apj, 848, 29
\bibitem[Dittmann et al.(2021)]{dit21}
Dittmann, A. J., Cantiello, M., Jermyn, A. S. 2021, \apj, 916, 48
\bibitem[Dokuchaev(1964)]{dok64}
Dokuchaev, V. P. 1964, Soviet Astronomy, 8, 23
\bibitem[Edgar(2004)]{edg04}
Edgar, R. G., 2004, NewAR, 48, 843
\bibitem[Hanuschik(1996)]{han96}
Hanuschik, R. W. 1996, \aap, 308, 170
\bibitem[Horedt(2000)]{hor00}
Horedt, G. P. 2000, \apj, 541, 821
\bibitem[Hoyle \& Lyttleton(1939)]{hoy39}
Hoyle, F., \& Lyttleton, R. A. 1939, PCPS, 35, 405
\bibitem[Ida et al.(2020)]{ida20}
Ida, S., Muto, T., Matsumura, S., Brasser, R. 2020, \mnras, 494, 5666
\bibitem[Igumenshchev \& Narayan(2002)]{igu02}
Igumenshchev, I. V., Narayan, R. 2002, \apj, 566, 137
\bibitem[Just \& Kegel(1990)]{jus90}
Just, A., Kegel, W. H. 1990, \aap, 232, 447
\bibitem[Kaaz et al.(2019)]{kaa19}
Kaaz, N., Antoni, A., Ram\'{\i}rez-Ruiz, E. 2019, \apj, 876, 142 
\bibitem[Krumholz et al.(2006)]{kru06}
Krumholz, M. R., McKee, C. F., Klein, R. I. 2006, \apj, 638, 369
\bibitem[Lee et al.(2014)]{lee14}
Lee, A. T., Cunningham, A. J., McKee, C. F., Klein, R. I. 2014, \apj,
783, 50 
\bibitem[Lescaudron et al.(2023)]{les23}
Lescaudron, S., Dubois, Y., Beckmann, R. S., Volonteri, M. 2023, \aap, 674, A217
\bibitem[Li et al.(2023)]{li23}
Li, Y.-P., Chen, Y.-X., Lin, D. N. C. 2023, \mnras, 526, 5346
\bibitem[Li et al.(2021)]{li21}
Li, Q.-C., Yang, Y.-P., Wang, F. Y., Xu, K., Shao, Y., Liu, Z.-N., Dai, Z.-G. 2021,
\apjl, 918, L5
\bibitem[Lin \& Murray(2007)]{lin07}
Lin, D. N. C., Murray, S. D. 2007, \apj, 661, 779
\bibitem[Livio et al.(1986)]{liv86}
Livio, M., Soker, N., de Kool, M., Savonije, G. J. 1986, \mnras, 222, 235
\bibitem[Lyttleton(1972)]{lyt72}
Lyttleton, R. A. 1972, \mnras, 160, 255
\bibitem[MacLeod \& Ram\'{\i}rez-Ruiz(2015)]{mac15}
MacLeod, M., Ram\'{\i}rez-Ruiz, E. 2015, \apj, 803, 41
\bibitem[Martin et al.(2025)]{mar25}
Martin, R. G., Lubow, S. H., Vallet, D., Overton, M., Lepp, S., Zhu, Z. 2025, \mnras, 539, L31
\bibitem[Matsuda et al.(1987)]{mat87}
Matsuda, T., Inoue, M., Sawada, K. 1987, \mnras, 227, 785
\bibitem[Matsuda et al.(1991)]{mat91}
Matsuda, T., Sekino, N., Sawada, K., Livio, M., Anzer, U., Borner, G. 1991, \aap, 248, 301
\bibitem[Matsuda et al.(2015)]{mat15}
Matsuda, T., Isaka, H., Ohsugi, Y. 2015, Prog. Theor. Exp. Phys., 11, 113E01
\bibitem[Mayer et al.(2007)]{may07}
Mayer, L., Kazantzidis, S., Madau, P., Colpi, M., Quinn, T., Wadsley, J. 2007, Science,
316, 1874
\bibitem[Muto et al.(2011)]{mut11}
Muto, T., Takeuchi, T., Ida, S. 2011, \apj, 737, 37
\bibitem[Negueruela \& Okazaki(2001)]{neg01}
Negueruela, I., Okazaki, A. T. 2001, \aap, 369, 108
\bibitem[Negueruela et al.(2001)]{neg01b}
Negueruela, I., Okazaki, A. T., Fabregat, J., Coe, M. J., Munari, U., Tomov, T. 2001, \aap,
369, 117
\bibitem[Okasaki et al.(2002)]{oka02}
Okazaki, A. T., Bate, M. R., Ogilvie, G. I., Pringle, J. E. 2002, \mnras, 337, 967
\bibitem[Okazaki et al.(2013)]{oka13}
Okazaki, A. T., Hayasaki, K., Moritani, Y. 2013, \pasj, 65, 41
\bibitem[Ostriker(1999)]{ost99}
Ostriker, E. C. 1999, \apj, 513, 252 
\bibitem[Papaloizou(2002)]{pap02}
Papaloizou, J. C. B. 2002, \aap, 388, 615
\bibitem[Papaloizou \& Larwood(2000)]{pap00}
Papaloizou, J, C. B., Larwood, J. D. 2000, \mnras, 315, 823
\bibitem[Proga \& Begelman(2003)]{pro03}
Proga, D., Begelman M. C. 2003, \apj, 582, 69
\bibitem[Prust \& Bildsten(2024)]{pru24}
Prust, L. J., Bildsten, L. 2024, \apj, 966, 103
\bibitem[Raga et al.(2022)]{rag22}
Raga, A. C., Cant\'{o}, J., Castellanos-Ram\'{\i}rez, A., Rodr\'{\i}guez-Gonz\'alez, A.,
Rivera-Ortiz, P. R. 2022, Rev. Mex. Astr. Astrop., 58, 215
\bibitem[Rein(2012)]{rei12}
Rein, H. 2012, \mnras, 422, 3611
\bibitem[Rephaeli \& Salpeter(1980)]{rep80}
Rephaeli, Y., Salpeter, E. E. 1980, \apj, 240, 20
\bibitem[Rosenthal et al.(2020)]{ros20}
Rosenthal, M. M., Chiang, E. I., Ginzburg, S., Murray-Clay, R. A. 2020, \mnras, 498, 2054
\bibitem[Ruffert(1996)]{ruf96}
Ruffert, M. 1996, \aap, 311, 817
\bibitem[S\'anchez-Salcedo(2012)]{san12}
S\'anchez-Salcedo, F. J. 2012, \apj, 745, 135
\bibitem[S\'anchez-Salcedo(2019)]{san19}
S\'anchez-Salcedo, F. J. 2019, \apj, 885, 152 
\bibitem[S\'anchez-Salcedo(2020)]{san20}
S\'anchez-Salcedo, F. J. 2020, \apj, 897, 142
\bibitem[S\'anchez-Salcedo et al.(2018)]{san18}
S\'anchez-Salcedo, F. J., Santill\'an, A., Chametla, R. O. 2018, \apj, 860, 129
\bibitem[S\'anchez-Salcedo \& Santill\'an(2025)]{san25}
S\'anchez-Salcedo, F. J., Santill\'an, A. 2025, \mnras, 537, 2647
\bibitem[Sawada et al.(1989)]{saw89}
Sawada, K., Matsuda, T., Anzer, U., Borner, G., Livio, M. 1989, \aap, 221, 263
\bibitem[Secunda et al.(2021)]{sec21}
Secunda, A., Hernandez, B., Goodman, J., Leigh, N. W. C., McKernan, B., Ford, K. E. S., Adorno, J. I. 2021, \apj, 908, 27
\bibitem[Soker(1990)]{sok90}
Soker, N. 1990, \apj, 358, 545
\bibitem[Suzuguchi et al.(2024)]{suz24}
Suzuguchi, T., Sugimura, K., Hosokawa, T., Matsumoto, T. 2024, \apj, 966, 7
\bibitem[Taam \& Fryxell(1988)]{taa88}
Taam, R. E., Fryxell, B. A. 1988, \apjl, 327, L73
\bibitem[Thun et al.(2016)]{thu16}
Thun, D., Kuiper, R., Schmidt, F., Kley, W. 2016, \aap, 589, A10
\bibitem[Toropina et al.(2012)]{tor12}
Toropina, O. D.; Romanova, M. M.; Lovelace, R. V. E. 2012, \mnras, 420, 810
\bibitem[Vicente et al.(2019)]{vic19}
Vicente, R., Cardoso, V., Zilh\~{a}o, M. 2019, \mnras, 489, 5424
\bibitem[Xiang-Gruess \& Papaloizou(2013)]{xia13}
Xiang-Gruess, M.,  Papaloizou, J. C. B. 2013, \mnras, 431, 1320
\bibitem[Xu \& Stone(2019)]{xu19}
Xu, W., Stone, J. 2019, \mnras, 488, 5162
\bibitem[Zhou et al.(2024)]{zho24}
Zhou, S., Sun, M., Liu, T., Wang, J.-M., Wang, J.-X., Xue, Y. 2024, \apjl, 966, L9

\end{thebibliography}
\end{document}